\documentclass[aps,PRB,twocolumn,superscriptaddress,preprintnumbers]{revtex4-2}

\usepackage{graphicx}
\usepackage{dcolumn}
\usepackage{bm}
\usepackage{amsmath}
\usepackage{braket}
\usepackage{here}
\usepackage{color}
\usepackage{amsfonts}
\usepackage{comment}

\begin{document}

\title{Topological Corner States in Bilayer and Trilayer Systems with \\ 
Vertically Stacked Topological Heterostructures}

\author{Natsuko Ishida}
\email{n-ishida@iis.u-tokyo.ac.jp} 
\address{Research Center for Advanced Science and Technology, The University of Tokyo, 4-6-1 Komaba, Meguro-ku, Tokyo 153-8505, Japan}
\author{Motohiko Ezawa} 
\address{Department of Applied Physics, The University of Tokyo, 7-3-1 Hongo, Bunkyo-ku, Tokyo 113-8656, Japan}
\author{Guangtai Lu}
\address{Research Center for Advanced Science and Technology, The University of Tokyo, 4-6-1 Komaba, Meguro-ku, Tokyo 153-8505, Japan}
\author{Wenbo Lin}
\address{Institute of Integrated Research, Institute of Science Tokyo, 2-12-1 Ookayama, Meguro-ku, Tokyo 152-8550, Japan}
\author{Yasutomo Ota}
\address{Department of Applied Physics and Physico-Informatics, Keio University, 3-14-1 Hiyoshi, Kohoku-ku, Yokohama, Kanagawa 223-8522, Japan}
\author{Yasuhiko Arakawa}
\address{Institute for Nano Quantum Information Electronics, The University of Tokyo, 4-6-1 Komaba, Meguro-ku, Tokyo 153-8505, Japan}
\author{Satoshi Iwamoto}
\address{Research Center for Advanced Science and Technology, The University of Tokyo, 4-6-1 Komaba, Meguro-ku, Tokyo 153-8505, Japan}
\address{Institute of Industrial Science, The University of Tokyo, 4-6-1 Komaba, Meguro-ku, Tokyo 153-8505, Japan}

\date{\today}

\begin{abstract}
We investigate bilayer and trilayer systems composed of topologically distinct, vertically stacked layers, forming topological heterostructures based on the Benalcazar-Bernevig-Hughes model. 
We find that a topological phase transition induced by interlayer coupling significantly alters the number of corner states in these topological structures. 
Furthermore, we find that traditional nested Wilson loop analysis inaccurately classifies certain phases, leading us to evaluate multipole chiral numbers (MCNs) as a more appropriate topological invariant for this scenario. 
The MCNs not only enable accurate classification of topological phases but also directly correspond to the number of zero-energy corner states, effectively characterizing $\mathbb{Z}$-class HOTI phases.
Our study proposes the novel concept of topological heterostructures, providing critical insights into the control of localized corner states within multilayer systems and expanding potential research directions.
\end{abstract}

\maketitle


\section{Introduction}
Topological insulators have emerged as fascinating materials in condensed matter physics\cite{Review2010,TopoReview2011,Review2016}, characterized by an insulating bulk and conductive edge states. These edge states arise from the topological features in the bulk band structure, which result in a non-zero topological invariant, thus demonstrating the concept of bulk-boundary correspondence\cite{HatsugaiPRL1993,Hatsugai1993}. The basic principles of topological materials have been extended beyond condensed matter systems to classical wave phenomena, such as photonics\cite{Ozawa2019,Khanikaev2013,Lu2014,Ota2020,Iwamoto2021}, acoustic waves\cite{Baile2015,He2016,Peng2016,Xue2022}, mechanical waves\cite{Kane2014,Huber2015,Huber2016}, thermal waves\cite{thermal2022}, and electric circuits\cite{Lee2018,Li2018}.
This expansion has led to attractive applications such as topological lasers\cite{Segev2018,Khajavikhan2018} and topological electronic devices\cite{Gilbert2021,Liu2022}.

In recent years, the concept of higher-order topological insulators (HOTIs)\cite{BBH2017,BBH-PRB2017} has gained significant attention, expanding the scope of topological materials. HOTIs are characterized by topological states that are at least two dimensions lower than that of the system itself; for example, a two-dimensional (2D) HOTI can host zero-dimensional corner states.

Building on this framework, the majority of research has been directed towards the study of 2D HOTIs\cite{Ezawa2018,EzawaKagome2018,Ezawa2018_2,Xie2018,Ezawa2019,CornerSlab2019,Li2020,Chen2021,Wang2021} and their three-dimensional vertically stacked extensions, which can host hinge states\cite{Taylor2018,Schindler2018,Sander2018,3Dreview2018,Taylor2020,Suotang2021}. 
These studies have aimed to uncover the unique topological properties of these systems and explore their potential applications in various fields, such as nanocavity \cite{Ota2019,Smirnova2020}, topological nanolasers\cite{Zhang2020,Kim2020}, nonlinear optics\cite{Kirsch2021,Ezawa2022}, and cavity quantum electrodynamics (QED)\cite{Xie2020,Kim2021}.

For multilayer systems, those composed of vertically stacked layers, known as van der Waals heterostructures \cite{VanderWaals2013}, have been extensively studied. 
These systems, characterized by weak interlayer coupling relative to in-plane coupling, have demonstrated unique topological properties\cite{Kezilebieke2020,Zhenhua2022} and show promise for applications in various fields\cite{CastellanosGomez2022,Qi_2023}. 
Despite this progress, the investigation of multilayer systems consisting of topologically distinct layers remains largely unexplored.

In this paper, we investigate multilayer systems composed of topologically distinct layers, forming topological heterostructures. We focus on the effects of corner states and topological phases in relation to interlayer coupling strength.
Our work is based on the Benalcazar-Bernevig-Hughes (BBH) model \cite{BBH2017}, the simplest model for studying quadrupole topological insulators. This model consists of $\pi$-flux square lattices and supports topological corner states.
The realization of quadrupole insulators has been demonstrated through various experimental setups, including optical ring resonator arrays \cite{Mittal2019}, waveguide arrays\cite{Schulz2022}, 
electric circuits\cite{Imhof2018}, acoustic system\cite{Serra2018}, and microwave resonator\cite{Peterson2018}.

Our findings indicate a topological phase transition induced by the interlayer coupling, which significantly influences the emergence and annihilation of topological corner states. 
Interestingly, we identify a phase in which the conventional nested Wilson loop method incorrectly classifies the topological phase as trivial, despite the presence of corner states.
Therefore, we calculate an alternative topological invariant, recently developed multipole chiral numbers (MCNs) \cite{Benalcazar2022}, to evaluate the topological phases.
Due to its ability to handle $\mathbb{Z}$-class chiral-symmetric HOTI systems\cite{Vaidya2023,Shi2024}, the MCN has been increasingly applied to various models \cite{Yun2023,Suotang2023,Acoustic2023,Yang2023,Yuhua2024,Chun2024}. 
We found that the MCN precisely identifies the phases present in our model and accurately matches the number of corner states at zero energy.
By adopting the MCN as a topological invariant, we not only identify changes in the number of corner states, but also uncover the topological phase transition induced by interlayer coupling.

\section{Topological corner states}
The BBH model consists of four sites in a unit cell where the $\pi$-flux is inserted by introducing the negative coupling in a plaquette, as shown with the red bond in Fig.~1 (a). 
There are two topologically distinct phases, determined by the ratio of the intracell coupling to the intercell coupling.
When the intra-cell coupling is weaker than inter-cell coupling, there exists four degenerate states that form a single set of corner modes, localized at each corner of the structure within a band gap at energy $E=0$.
The bulk Hamiltonian for the system shown in Fig.~1 (a) is given by
\begin{figure}[H]
\includegraphics[]{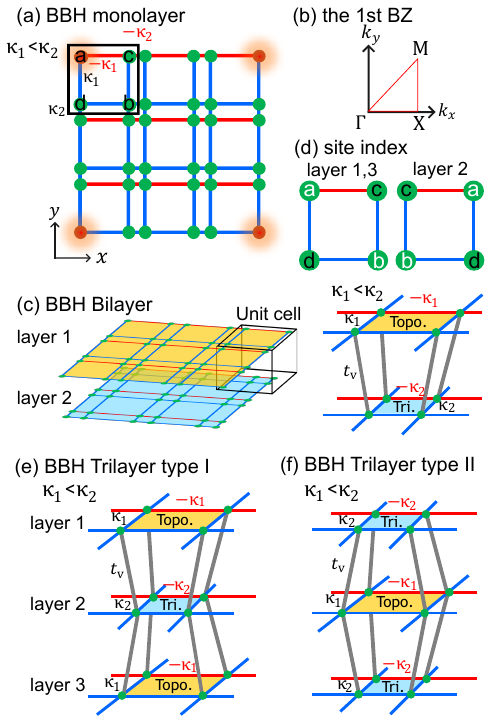}
\caption{\label{fig:configurations} (a) Blue and red bonds represent the positive and negative couplings, respectively. Corner states, shown in red at each corner, appear when $\kappa_1 < \kappa_2$. 
(b) the 1st Brillouin zone. 
(C) Illustration of the BBH bilayer system composed of topological and trivial layers.
$\kappa_1$ ($\kappa_2$) and $\kappa_2$ ($\kappa_1$) denote the intra- and inter-cell couplings for topological (trivial) system, respectively.
(d) The labeling lattice sites a, b, c, and d represent the Hamiltonian basis for each layer.
(e,f) Two distinct BBH trilayer systems in which topologically distinct layers are stacked on top of each other.}
\end{figure}
\begin{widetext}    
\begin{equation}
\mathcal{H}_{\mathrm{BBH}}(k_x,k_y)
=
\begin{pmatrix}
0 & 0 & -(\kappa_1+\kappa_2 e^{-i k_x})  &  \kappa_1+\kappa_2 e^{-i k_y}\\
0 & 0 & \kappa_1+\kappa_2 e^{i k_y} & \kappa_1+\kappa_2 e^{i k_x}\\
-(\kappa_1+\kappa_2 e^{i k_x}) & \kappa_1+\kappa_2 e^{-i k_y} & 0 & 0\\
\kappa_1+\kappa_2 e^{i k_y} & \kappa_1+\kappa_2 e^{-i k_x} & 0 & 0 \\
\end{pmatrix}.
\end{equation}
\end{widetext}
Throughout this paper, we use $\kappa_1$ ($\kappa_2$) and $\kappa_2$ ($\kappa_1$) as intracell and intercell couplings for the topological (trivial) system, respectively, with its corresponding Hamiltonian $\mathcal{H}_{\text{BBH}}^{\text{Topo}}\left( k_{x},k_{y}\right)$ ($\mathcal{H}_{\text{BBH}}^{\text{Tri}}\left( k_{x},k_{y}\right)$).
Here, we suppose $\kappa_1 < \kappa_2$, and both are positive.

The BBH system has two occupied degenerate bands that possess two canceling dipole moments and a non-vanishing quadrupole moment. 
By adopting the Wilson loop calculation, the degeneracy is lifted, resulting in two distinct single-band subspaces where Wannier bands become non-degenerate. 
Consequently, each Wannier band carries its own topological invariant, which can be obtained through the nested Wilson loop calculation \cite{BBH2017}. 
Following the calculation, a quadrupole moment $q_{xy}=\frac{1}{2}$ indicates that the system is topological, while $q_{xy}=0$ means that the system is trivial.

In this paper, we consider a vertically stacked configuration composed of topologically distinct layers, taking into account the vertical coupling of the nearest neighbor site denoted by $t_\mathrm{v}$. 
Fig.~\ref{fig:configurations} (c) illustrates the considered BBH bilayer system and its unit cell, where the top and bottom layers are topological and trivial, respectively. 
For trilayer systems, we consider two cases as shown in Figs.~\ref{fig:configurations} (e) and (f), where the arrangements of the topological and trivial layers are flipped.

\subsection{Topological corner states in BBH bilayer system}
\begin{figure}[t]
\includegraphics[]{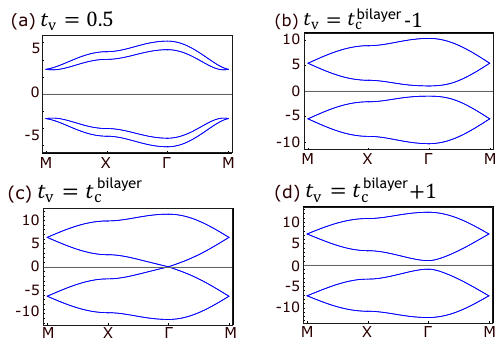}
\caption{
\label{fig:TopoTri_band} Interlayer coupling $t_\mathrm{v}$ dependence of band structures for BBH bilayer system where top and bottom layers are composed of two topologically distinct layers. Parameters are chosen as 
(a) $t_\mathrm{v}$=0.5,
(b) $t_\mathrm{v}=t_\mathrm{c}^{\rm{bilayer}}-1$,
(c) $t_\mathrm{v}$=$t_\mathrm{c}^{\rm{bilayer}}$ ,
(d) $t_\mathrm{v}=t_\mathrm{c}^{\rm{bilayer}}+1$, $\kappa_1$=1, and $\kappa_2$=3.
$t_\mathrm{c}^{\rm{bilayer}}(=4\sqrt{2}$) denotes the critical value where a band gap closes.}
\end{figure}

The Hamiltonian describing the bilayer system, shown in Fig.~\ref{fig:configurations} (c), is given by
\begin{equation}
\mathcal{H}_{\text{bilayer}}=
\left( 
\begin{array}{cc}
\mathcal{H}_{\text{BBH}}^{\text{Topo}} & \mathcal{H}_{\text{v}} \\ 
\mathcal{H}_{\text{v}} & \mathcal{H}_{\text{BBH}}^{\text{Tri}}
\end{array}
\right),
\end{equation}

where 
\begin{equation}
\mathcal{H}_{\text{v}}=t_{\text{v}}\left( 
\begin{array}{cccc}
0 & 0 & 1 & 0 \\ 
0 & 0 & 0 & 1 \\ 
1 & 0 & 0 & 0 \\ 
0 & 1 & 0 & 0
\end{array}
\right)
=
t_{\text{v}}
\sigma_x \otimes \mathbb{I}_2.
\end{equation}
Here, we use the site index shown in Fig.~1(d) in which chiral symmetry is preserved. This ordering is essential for the theoretical analysis presented later in this paper.
$\sigma_x$ and $\mathbb{I}_2$ are the Pauli matrix and two-by-two identity matrix, respectively.
The band structure has four pairs of doubly degenerate bands, and exhibits both opening and closing of a band gap around the $\Gamma$ point as $t_\mathrm{v}$ changes.
(Fig.~\ref{fig:TopoTri_band}).
The critical value $t_\mathrm{c}^{\rm{bilayer}}$ corresponds to the point at which the band gap closes, defined as $t_\mathrm{c}^{\rm{bilayer}}=\sqrt{2}(\kappa_1+\kappa_2)$. 
See Appendix A for details.

\begin{figure}[t]
\includegraphics[]{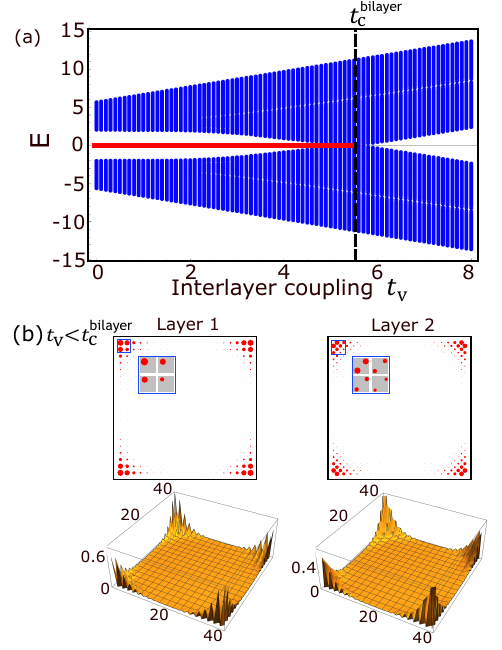}
\caption{\label{fig:corner_bilayer}
(a) Energy spectrum vs $t_\mathrm{v}$ for BBH bilayer system. 
Blue dots represent the bulk spectrum, with or without gapped edges, whereas the red line corresponds to a single set of corner modes. The gapped edge states appear when corner modes exist.
(b) Field amplitude summed over the four corner modes at $E=0$ for $t_\mathrm{v}$=$t_\mathrm{c}^{\rm{bilayer}}-1$.
The inset shows an enlarged view of a single corner, with the gray area representing a unit cell.
Parameters are chosen as $\kappa_1$=1, $\kappa_2=3$ and $N=20$.}
\end{figure}

We now consider a finite BBH model with square geometry, with each layer forming a BBH lattice consisting of $2N\times 2N$ sites.
Fig.~\ref{fig:corner_bilayer}(a) shows the energy spectrum as a function of $t_\mathrm{v}$. 
In the weak interlayer coupling regime ($t_\mathrm{v}<t_\mathrm{c}^{\rm{bilayer}}$), the system possesses a single set of corner modes, consisting of a total of four degenerate states, each predominantly localized at the corners of the layers.
This is depicted by the red line in the figure.
Fig.~\ref{fig:corner_bilayer}(b) shows the field amplitude of the corner state, summed for all four states. 
An enlarged view of the corner modes reveals their localization on a single sublattice at each corner, attributed to chiral symmetry. 
A detailed analysis of the field distribution, considering the limit $\kappa_1=0$, is presented in Appendix B.
These corner states disappear when the interlayer coupling exceeds the critical value
$t_\mathrm{c}^{\rm{bilayer}}$.
This disappearance can be understood through perturbation theory, as explained in detail in Appendix D.
Sudden disappearance of the corner states, accompanied by a gap closing, suggests a topological phase transition associated with increasing interlayer coupling strengths.

To determine the topological phases, we calculate the quadrupole moment $q_{xy}$ \cite{BBH2017} using the nested Wilson loop.
As a result, we find that $q_{xy}=\frac{1}{2}$ for $t_\mathrm{v}<t_\mathrm{c}^{\rm{bilayer}}$, identifying the system as topological, while $q_{xy}=0$ for $t_\mathrm{v}>t_\mathrm{c}^{\rm{bilayer}}$, indicating a trivial phase.
These results indicate that the interlayer coupling induces a topological phase transition in the BBH bilayer system.
The presence and absence of topological corner states changes at the critical value $t_\mathrm{c}^{\rm{bilayer}}$ as the band gap closes and re-opens.
This critical value marks the transition point where the number of topological corner states changes.
The detailed calculation of the quadrupole moment is shown in Appendix C.

\subsection{Topological corner states in BBH trilayer system type $\mathrm{I}$ (topological-trivial-topological structure)}
We proceed to study a BBH trilayer system depicted in Fig.~1(e).
This arrangement consists of a trivial layer positioned between two topological layers, where the layers are coupled by interlayer coupling $t_\mathrm{v}$. 

The system Hamiltonian is given by
\begin{equation}
\mathcal{H}_{\text{trilayerI}}=
\left( 
\begin{array}{ccc}
\mathcal{H}_{\text{BBH}}^{\text{Topo}} & \mathcal{H}_{\text{v}} & 0 \\ 
\mathcal{H}_{\text{v}} & \mathcal{H}_{\text{BBH}}^{\text{Tri}} & \mathcal{H}_{\text{v}} \\ 
0 & \mathcal{H}_{\text{v}} & \mathcal{H}_{\text{BBH}}^{\text{Topo}}
\end{array}
\right),
\end{equation}
with the interlayer coupling matrix $\mathcal{H}_{\text{v}}$.
Similarly to the bilayer system, the labeling of sites in the Hamiltonian varies between odd and even layers to maintain the chiral symmetry, as shown in Fig.~1(d).
This labeling is crucial for applying the methods we will discuss in the next section.
The trilayer system has six pairs of doubly degenerate bands.
Similarly to the bilayer system, the band gap closes at a critical value $t_\mathrm{c}^{\rm{trilayer}}=\kappa_1+\kappa_2$ (see Appendix A). The band gap reopens again when $t_\mathrm{v}$ exceeds the critical value. 
\begin{figure}[t]
\includegraphics[]{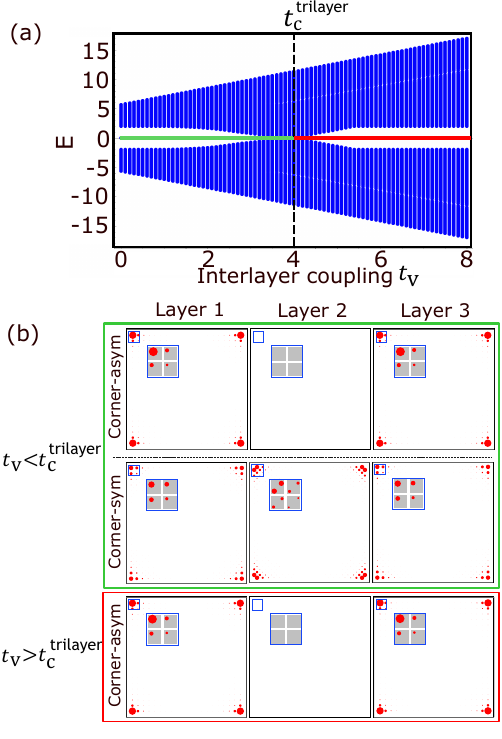}
\caption{\label{fig:corner_trilayer1}(a) Energy spectrum vs $t_\mathrm{v}$ for BBH trilayer system type $\mathrm{I}$ (topological-trivial-topological structure). 
Blue dots represent the gapped edge states and bulk spectrum. The green and red lines correspond to two and a single set of corner modes, respectively.
(b) Field amplitude summed over the four corner modes at $E=0$ for $t_\mathrm{v}$=$t_\mathrm{c}^{\rm{trilayer}}-1$ and $t_\mathrm{v}$=$t_\mathrm{c}^{\rm{trilayer}}+1$.
'Corner-asym' persists even after exceeding the critical value, while 'Corner-sym' vanishes at the critical value $t_\mathrm{c}^{\rm{trilayer}}$.
The inset shows an enlarged view of a single corner, with the gray area representing a unit cell.
Parameters are chosen as $\kappa_1$=1, $\kappa_2=3$, and $N=20$.}
\end{figure}

For a finite BBH trilayer system with square geometry, with each layer consisting of $2N \times 2N$ sites, we calculate the eigenenergies as a function of the interlayer coupling strength(Fig~\ref{fig:corner_trilayer1}(a)).
When $t_\mathrm{v}<t_\mathrm{c}^{\rm{trilayer}}$, which is below the critical value, there are two sets of corner states, with a total of eight (illustrated by the green line in the figure).
In contrast, when $t_\mathrm{v}$ exceeds $t_\mathrm{c}^{\rm{trilayer}}$, only a single set of corner states survives, with a total number of four (indicated by the red line).
Each of these sets consists of four degenerate modes at $E=0$. 
These two sets of corner states are distinguished by their unique field distributions(Fig.~\ref{fig:corner_trilayer1} (b)). 
Specifically, the corner states that disappear at the critical interlayer coupling show non-zero field amplitudes across all layers, 
which corresponds to symmetric coupling between the corners (Corner-sym).
Conversely, the corner states that remain after exceeding the critical interlayer coupling exhibit a complete absence of field amplitude in the middle layer, which corresponds to anti-symmetric coupling between the corners (Corner-asym).
The detailed analysis of the field distribution for the corner states under the $\kappa_1=0$ limit is provided in Appendix B.
At large interlayer couplings, the presence of topological corner states, 'Corner-asym', is explained through an effective model near zero energy, derived based on the perturbation theory. See Appendix D.

As discussed in subsection \(\mathrm{II.A}\) of the bilayer system, we calculate the quadrupole moment $q_{xy}$ to identify the topological phases of the trilayer system.
Consequently, our analysis shows that $q_{xy}=0$ for $t_\mathrm{v}<t_\mathrm{c}^{\rm{trilayer}}$, classifying the system as trivial, despite the presence of two sets of corner states at $E=0$.
In the case where $t_\mathrm{v}>t_\mathrm{c}^{\rm{trilayer}}$, we have $q_{xy}=\frac{1}{2}$, indicating the system as topological.
This leads to the issue where the quadrupole moment fails to fully capture the topological phases in certain instances.
Therefore, the introduction of another invariant is essential to accurately describe the topological phases of both BBH bilayer and trilayer systems.
We will calculate and discuss this alternative invariant in the next section.

\subsection{Topological corner states in BBH trilayer system type $\mathrm{II}$ (trivial-topological-trivial structure)}
Next, we study a trilayer system that involves a topological layer sandwiched between two trivial layers, each layer coupled by an interlayer coupling $t_\mathrm{v}$. 
This arrangement is depicted in Fig.~\ref{fig:configurations} (f).
In this case, the Hamiltonian is given by 
\begin{equation}
\mathcal{H}_{\text{trilayerI\hspace{-1.2pt}I}}=
\left( 
\begin{array}{ccc}
\mathcal{H}_{\text{BBH}}^{\text{Tri}} & \mathcal{H}_{\text{v}} & 0 \\ 
\mathcal{H}_{\text{v}} & \mathcal{H}_{\text{BBH}}^{\text{Topo}} & \mathcal{H}_{\text{v}} \\ 
0 & \mathcal{H}_{\text{v}} & \mathcal{H}_{\text{BBH}}^{\text{Tri}}
\end{array}
\right).
\end{equation}
Its critical value is determined by $t_\mathrm{c}^{\rm{trilayer}}=\kappa_1+\kappa_2$, which is identical to that in the previous trilayer case.
\begin{figure}[t]
\includegraphics[]{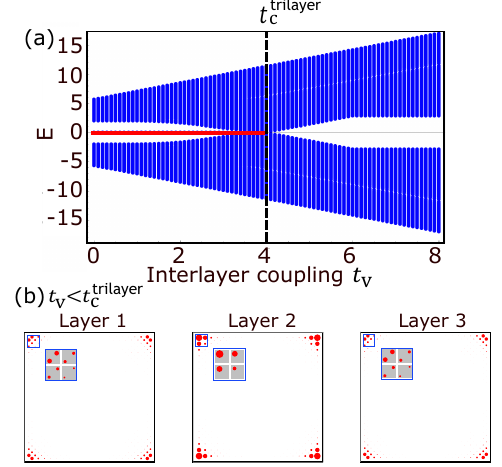}
\caption{\label{fig:corner_trilayer2}
(a) Energy spectrum vs $t_\mathrm{v}$ for BBH trilayer system type $\mathrm{II}$ (trivial-topological-trivial structure). 
Blue dots correspond to the bulk spectrum, with or without gapped edges, whereas the red line represents a single set of corner modes. Gapped edges are present when the system exhibits corner states.
(b) Field amplitude summed over the four corner modes at $E=0$ for $t_\mathrm{v}$=$t_\mathrm{c}^{\rm{trilayer}}-1$.
The inset shows an enlarged view of a single corner, with the gray area representing a unit cell.
Parameters are chosen as $\kappa_1$=3, $\kappa_2=1$ and $N=20$.}
\end{figure}

Fig.~\ref{fig:corner_trilayer2}(a) shows the energy spectrum as a function of $t_\mathrm{v}$. 
There is a single set of corner modes that includes four degenerate corner states at $E=0$ (depicted by the red line) when $t_\mathrm{v}<t_\mathrm{c}^{\rm{trilayer}}$; however, all corner modes disappear above the critical value $t_\mathrm{c}^{\rm{trilayer}}$.
An enlarged view of a single mode reveals that the corner mode distribution is localized on the sublattice opposite to that of the topological-trivial-topological structure. See Appendix B for details.
The absence of topological corner states at large interlayer couplings can be explained by an effective model near zero energy, derived through perturbation theory, as detailed in Appendix D.

The field amplitudes of the corner modes, shown in Fig.~\ref{fig:corner_trilayer2} (b), are distributed across each layer.
The quadrupole moments are obtained as $q_{xy}=\frac{1}{2}$ for $t_\mathrm{v}<t_\mathrm{c}^{\rm{trilayer}}$, identifying the phase as topological, while $q_{xy}=0$ for $t_\mathrm{v}>t_\mathrm{c}^{\rm{trilayer}}$, classifying it as trivial.
We note that in BBH trilayer systems with large interlayer coupling $t_\mathrm{v}$, the size of the band-gap varies between two distinct systems, despite the fact that they have identical band structures, as shown in Figs. 4(a) and 5(a). 
This discrepancy is caused by the presence of gapped edge states within the band gap, as illustrated in Fig.~4(a).

\section{2D winding number in real space}
As we observed a discrepancy between the quadrupole moment and the presence of corner states when $t_\mathrm{v}<t_\mathrm{c}^{\rm{trilayer}}$, in the trilayer case ('topological-trivial-topological' structure), we need to calculate another topological invariant to discuss its topological phase in detail. 
We choose the MCN as the additional invariant, which is capable of accurately evaluating the topological phases of $\mathbb{Z}$-class chiral-symmetric  HOTI systems, as recently introduced by \cite{Benalcazar2022}.
The MCNs are analogous to the 2D winding number in real space, as established in \cite{Lin2021} and recently extended in \cite{Lin2024}.
This development is especially crucial in situations where existing theoretical frameworks, such as the nested Wilson loop, may erroneously categorize certain phases as trivial.
The MCN provides an innovative approach for analyzing these systems. 
Notably, the value of MCN directly correlates with the number of degenerate zero-energy corner states present in a finite system, thereby offering a more precise invariant for examining these unique topological states.
In this study, we use the MCNs to identify the topological phases of both BBH bilayer and trilayer systems in their chiral-symmetric forms.
This is particularly relevant in cases where the conventional nested Wilson approach inaccurately classified the phase as trivial, despite the presence of zero-energy corner states.

To compute the topological invariant for the BBH bilayer system, we modify the Hamiltonian in its chiral-symmetric form for its finite-size system
as follows:
\begin{equation}
    \mathcal{H}_{\text{bilayer}}=
    \begin{pmatrix}
0 & h_{\text{bilayer}}\\
h^{\dagger}_{\text{bilayer}} & 0
\end{pmatrix}.
\end{equation}
Note that the labeling of sites in the Hamiltonian differs between layer 1 and layer 2 in order to preserve the chiral symmetry, as shown in Fig.~1(d). In addition, sites a and b belong to sublattice A, while sites c and d belong to sublattice B.
When a 2D square lattice has $N$ unit cells in the $x$ and $y$-directions (i.e., each layer consists of $N \times N$ unit cells), $h_{\text{bilayer}}$ is a $4N^2 \times 4N^2$ matrix. 
Here, the eigenstates of $\mathcal{H}_{\text{bilayer}}$ in equation (6) are denoted as
\begin{widetext}
$\ket{\psi}^T=(a_1^1,\cdots,a_{N^2}^1,b_1^1,\cdots,b_{N^2}^1,a_1^2,\cdots, a_{N^2}^2,b_1^2,\cdots,b_{N^2}^2,c_1^1,\cdots,c_{N^2}^1,d_1^1,\cdots,d_{N^2}^1,c_1^2,\cdots,c_{N^2}^2,d_1^2,\cdots,d_{N^2}^2)$. 
\end{widetext}    
The superscripts represent the layer index, where layer 1 and 2 correspond to the upper and lower layer, respectively.
The quadrupole moment operator for the BBH bilayer system is represented by
a following sublattice quadrupole moment operator\cite{Benalcazar2022,Wheeler2019}
\begin{equation}
    \mathcal{Q}_{xy}^S=\sum_{\mathbf{R},\alpha\in \mathcal{S}} 
    e^{
    -i \frac{2\pi xy}{N^2}}
    \ket{\mathbf{R},\alpha}
    \bra{\mathbf{R},\alpha}
\end{equation}
for $\mathcal{S}=A,B$.
Each layer consists of $N \times N$ unit cells, each labeled as $\mathbf{R}=(x,y)$.
We diagonalize the Hamiltonian $\mathcal{H}_{\text{bilayer}}$ by using the singular value decomposition (SVD) of $h_{\text{bilayer}}=U_A\Sigma U_B^{\dagger}$, where $U_A$ ($U_B$) denotes a $4N^2 \times 4N^2$ unitary matrix while $\Sigma$ is a diagonal matrix of singular values.
Using $\mathcal{\tilde{Q}}_{xy}^S=U_{\mathcal{S}}^{\dagger}  \mathcal{Q}_{xy}^S U_{\mathcal{S}}$, the MCN is defined as \cite{Benalcazar2022}
\begin{equation}
    N_{xy}=\frac{1}{2\pi i} \mathrm{Tr} \left[ \mathrm{log} (  \mathcal{\tilde{Q}}_{xy}^A \mathcal{\tilde{Q}}_{xy}^{B \dagger}) \right].
\end{equation}

Our calculations reveal that the MCN invariant is $N_{xy}=1$ for $t_\mathrm{v}<t_\mathrm{c}^{\rm{bilayer}}$ while $N_{xy}=0$ for $t_\mathrm{v}>t_\mathrm{c}^{\rm{bilayer}}$, where the MCN number matches the number of corner states at $E=0$. 
This result is consistent with the Wilson loop calculation.
The comprehensive calculations for MCNs are presented in Appendix F.
It is important to note that, to calculate the topological invariant, sufficiently large systems are required.

This calculation can be readily adopted for trilayer cases.
First, we rewrite the trilayer Hamiltonian in equation (4) and (5) into a form that preserves chiral symmetry, given by
\begin{equation}
    \mathcal{H}_{\text{trilayer}}=
    \begin{pmatrix}
0 & h_{\text{trilayer}}\\
h^{\dagger}_{\text{trilayer}} & 0
\end{pmatrix}.
\end{equation}

The Hamiltonian $h_{\text{trilayer}}$ is represented by a matrix of size $6N^2 \times 6N^2$.
By using the quadrupole moment operators in equation (7), we obtain the MCN, given by equation (8).
For a trilayer system depicted in Fig.~1(e), the invariant is obtained as $N_{xy}=2$ for $t_\mathrm{v}<t_\mathrm{c}^{\rm{trilayer}}$ while $N_{xy}=1$ for $t_\mathrm{v}>t_\mathrm{c}^{\rm{trilayer}}$.
The MCN number perfectly matches the number of corner states at $E=0$.
Similarly, for the alternative trilayer system illustrated in Fig.~1(f), we obtain $N_{xy}=1$ for $t_\mathrm{v}<t_\mathrm{c}^{\rm{trilayer}}$ and $N_{xy}=0$ for $t_\mathrm{v}>t_\mathrm{c}^{\rm{trilayer}}$, which again coincides with the number of corner states at $E=0$.
Therefore, we conclude that the MCN accurately captures the topological phases that even the quadrupole moment may fail to identify.

\section{Discussion and Conclusions}
We investigated the BBH bilayer and trilayer systems, constructed by stacking topologically distinct layers, forming what we refer to as topological heterostructures.
By adjusting the interlayer coupling strengths, we identified changes in the number of corner states, indicating a topological phase transition. 
This topological phase transition, accompanied by the closing of the band gap, is induced by the interlayer coupling.
\begin{figure}[t]
\includegraphics[]{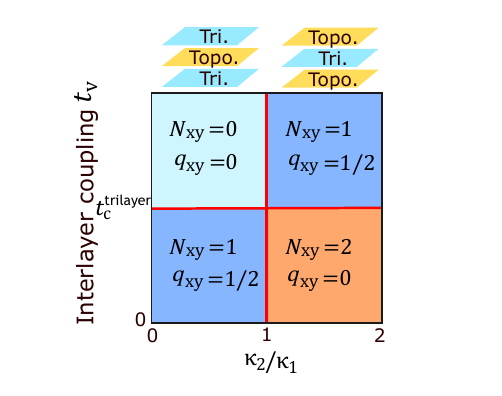}
\caption{\label{fig:phase_diagram}
Topological phase diagram of the BBH trilayer system as a function of
the the coupling strengths ratio and the interlayer coupling, highlighting areas defined by distinct multipole chiral numbers $N_{xy}$ and quadrupole moments $q_{xy}$. The closing of band gaps is marked by red solid lines.}
\end{figure}

To identify the topological phases within these topological heterostructures, we first calculated the quadrupole moments using the nested Wilson loop approach, a well-known method for analyzing quadrupole insulators. 
The quadrupole moments explained the topological phases of BBH bilayer and 'trivial-topological-trivial' trilayer structures.
However, we observed a mismatch between the quadrupole moments and the presence of corner states in the 'topological-trivial-topological' trilayer structure when the interlayer coupling was below the critical value.
In contrast, MCNs, recently reported as a topological invariant for $\mathbb{Z}$-class chiral-symmetric HOTI systems, can systematically explain the results.
We also confirmed that the number of corner states coincides with the MCNs across all interlayer coupling regimes for all three structures we considered. 

The categorization of the phase transition induced by interlayer coupling is as follows:
\begin{description}
   \item[Bilayer (topological-trivial)]\mbox{}\\
           from the topological phase with $N_{xy}=1$ to the trivial phase with $N_{xy}$=0.
   \item[Trilayer (topological-trivial-topological)]\mbox{}\\
	    from the topological phase with $N_{xy}=2$ to the topological phase with $N_{xy}=1$.
   \item[Trilayer (trivial-topological-trivial)]\mbox{}\\
           from the topological phase with $N_{xy}=1$ to the trivial phase with $N_{xy}$=0.
\end{description}
Figure 6 shows the topological phase diagram, which illustrates both the MCNs and the quadrupole moments in BBH trilayer systems.

As this topological phase transition is accompanied by the appearance and disappearance of corner states, these topological heterostructures may offer new possibilities for future applications. The vertically stacked layer configuration discussed in our paper can be implemented using existing platforms, such as electrical circuits. Additionally, the field distribution of the corner modes suggests that their intensity, which extends across multiple layers, can be further enhanced through layer stacking.
This finding indicates that topological heterostructures are potentially adoptable for the development of novel topological lasers when implemented on photonic platforms.
We believe that our study offers valuable insight into the control of localized corner states in multilayer systems, paving the way for new research directions in vertically stacked topological heterostructures.

\section*{Appendix A: bulk energy for the BBH bilayer and trilayer system}

($\mathrm{i}$)
The bulk spectrum of the bilayer system $\mathcal{H}_{\text{bilayer}}$ is obtained as
\begin{equation}
E= \pm \sqrt{2F+t_{\text{v}}^{2}\pm \sqrt{2G}},
\end{equation}
with 
\begin{eqnarray}
F &=&\kappa _{1}^{2}+\kappa _{2}^{2}+\kappa _{1}\kappa _{2}\left( \cos
k_{x}+\cos k_{y}\right) , \\
G &=&\left( \kappa _{1}+\kappa _{2}\right) ^{2}t_{\text{v}}^{2}\left( \cos
k_{x}+\cos k_{y}+2\right) ,
\end{eqnarray}
where each energy is two-fold degenerated. At the $\Gamma$ point, the energy is obtained as
\begin{eqnarray}
E=\pm \sqrt{2}|\kappa_{1}+\kappa_{2}|\pm t_{\text{v}}.
\end{eqnarray}
The gap closes at 
\begin{eqnarray}
t_{\text{v}}=\pm \sqrt{2}(\kappa_{1}+\kappa_{2}).
\end{eqnarray}

($\mathrm{ii}$) The bulk spectrum of the trilayer systems $\mathcal{H}_{\text{trilayerI}}$ and $\mathcal{H}_{\text{trilayerII}}$ is obtained as
\begin{equation}
E=\pm \sqrt{2}\sqrt{F},\qquad \pm \sqrt{2}\sqrt{F+t_{\text{v}}^{2}\pm \sqrt{G}},
\end{equation}
where each energy is two-fold degenerated. It is remarkable that the bulk
spectrum is identical between $\mathcal{H}_{\text{trilayerI}}$ and $\mathcal{H}_{\text{trilayerII}}$. At the $\Gamma$ point, the energy is obtained as
\begin{eqnarray}
E=\pm \sqrt{2}|\kappa_{1}+\kappa_{2}|, \qquad \pm \sqrt{2}(\kappa_{1}+\kappa_{2}\pm t_{\text{v}}).
\end{eqnarray}
The gap closes at 
\begin{eqnarray}
t_{\text{v}}=\pm (\kappa_{1}+\kappa_{2}).
\end{eqnarray}

\section*{Appendix B: The limit model $\kappa_1=0$}
Corner states that spread over two or three layers exhibit unique distributions in each layer. 
For example, in the case of a BBH bilayer system composed of topological and trivial, the corner cell within the corner unit cell in layer 1 has an amplitude, while in layer 2, the second cell from the corner exhibits an amplitude.
Each layer exhibits staggered finite amplitudes only on a single sublattice of the system, due to chiral symmetric protection \cite{Benalcazar2022}. The amplitudes at each site can be explained by analyzing the limit model with $\kappa_1=0$.

Figure 7(a) illustrates one corner of the BBH bilayer system in the $\kappa_1=0$ limit model. 
When a single corner mode exists in layer 1, the site amplitudes in layer 2 are given by $a=-b=C_1 \frac{t_{\text{v}}}{2\kappa_2}$, and $c=d=0$.
\begin{figure}[t]
\includegraphics[]{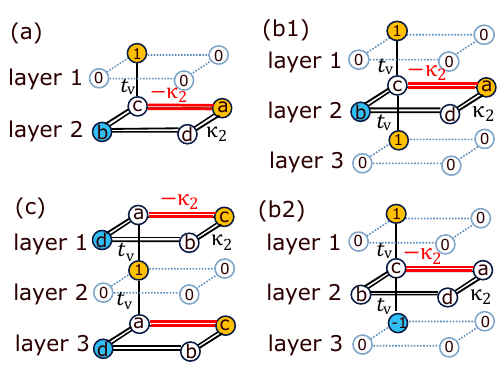}
\caption{\label{fig:limitmodel} The limit model $\kappa_1=0$ for (a) BBH bilayer,
(b1, b2) 'Topological-Trivial-Topological' BBH trilayer systems and 
(c) 'Trivial-Topological-Trivial' BBH trilayer systems.
(a, b1, c) Below the critical interlayer coupling strength ($t_\mathrm{v} < t_\mathrm{c}$).
(b2) Above the critical interlayer coupling strength ($t_\mathrm{v} > t_\mathrm{c}$).
}
\end{figure}

In the $\kappa_1=0$ limit model of the BBH trilayer (topological-trivial-topological structure), in-phase corner modes appear in layer 1 and layer 3 for cases below the critical interlayer coupling strength, as shown in Fig.~7(b1). This results in symmetrical coupling, where the site amplitudes in layer 2 are $a=-b=C_2 \frac{t_{\text{v}}}{\kappa_2}$, and $c=d=0$. In contrast, for cases above the critical interlayer coupling strength, out-of-phase corner modes emerge in layer 1 and layer 3, as illustrated in Fig.~7(b2). This results in the absence of amplitude in layer 2 due to anti-symmetric coupling.

In the $\kappa_1=0$ limit model of the BBH trilayer (trivial-topological-trivial structure), only a single corner mode exists in layer 2, which belongs to sublattice B. As a result, finite amplitudes appear in the B sublattices of layers 1 and 3, given by $c=-d=C_3 \frac{t_{\text{v}}}{\kappa_2}$, while $a=b=0$.
Here, $C_1, C_2, C_3$ are normalization constants.

\section*{Appendix C: Quadrupole moment $q_{xy}$ in the BBH bilayer and trilayer system}

In the BBH model, the bulk quadrupole moment $q_{xy}$, its boundary polarizations $p_{x,y}$ and corner charges $Q$ have the following relationship: $|p^{\mathrm{edge} \pm y}_x|=|p^{\mathrm{edge} \pm x}_y|=|Q^{\mathrm{corner} \pm x, \pm y}|=|q_{xy}|$ \cite{BBH2017}.
The requirement for quantizing the quadrupole moment $q_{xy}$ is the existence of three symmetries, which are the inversion symmetry and $x,y$-reflection symmetry under a gauge transformation.
These constraints lead to quantized values of the Wannier band polarizations $p_y^{v_x^{\pm}},p_x^{v_y^{\pm}}\overset{\mathcal{I},\mathcal{M}_x,\mathcal{M}_y}{=} 0$ or $1/2$.
Furthermore, when the system preserves $C_4$ symmetry, the classification of the Wannier bands is $\mathbb{Z}_2$ \cite{Taylor2018}.
Therefore, the quadrupole moment $q_{xy}$ can be described by
$q_{xy}=p_y^{v_x^{\pm}}=p_x^{v_y^{\pm}}=0$ or $1/2$ and the topological invariant is defined by the polarizations which are obtained as follows: 
\begin{equation}
p_y^{v_x^{\pm}} = -\frac{i}{2 \pi}\frac{1}{N_x}\sum_{k_x} \log[\Tilde{W}_{y,k_x}^{\pm}]= 0 \ \mathrm{or} \ \frac{1}{2}
\end{equation}
where $\Tilde{W}_{y,k_x}^{\pm}$ is a nested Wilson loop along $k_y$ \cite{BBH2017}. 
The quadrupole moment of $\frac{1}{2}$ and 0 implies that the system is topological and trivial, respectively.
\begin{figure}[t]
\includegraphics[]{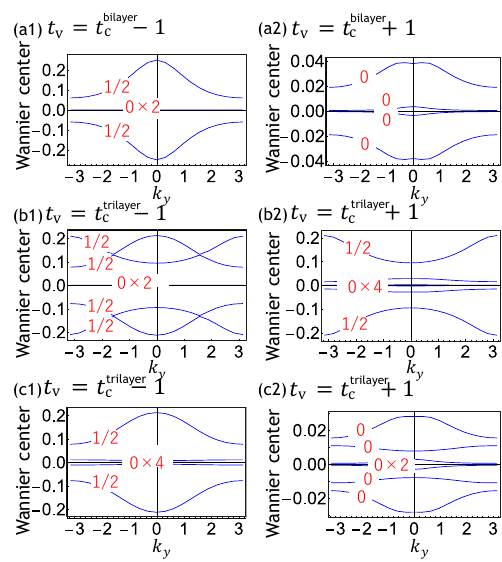}
\caption{\label{fig:Wannierband} Wannier bands of (a) BBH bilayer,
(b) 'Topological-Trivial-Topological' BBH trilayer systems and 
(c) 'Trivial-Topological-Trivial' BBH trilayer systems.
(a1)-(c1) Below the critical interlayer coupling strength ($t_\mathrm{v}$=$t_\mathrm{c}$-1).
(a2)-(c2) Above the critical interlayer coupling strength ($t_\mathrm{v}$=$t_\mathrm{c}$+1).
Each band is assigned a number that represents its polarization. Parameters are chosen as $\kappa_1$=1 and $\kappa_2=3$ for (a) and (b), whereas the values are reversed ($\kappa_1$=3 and $\kappa_2=1$) for (c).}
\end{figure}

By adopting the Wilson loop calculation to the occupied bands of the BBH bilayer shown in Fig.~2, we obtain four non-degenerate Wannier bands, as depicted in Fig.~\ref{fig:Wannierband}(a1,a2). 
The gapping of the Wannier bands results from the non-commutativity of the two reflection symmetries, $\mathcal{M}_x^{\mathrm{bi}}$ and $\mathcal{M}_y^{\mathrm{bi}}$, as discussed in \cite{BBH2017}. 
These two reflection symmetries in the bilayer system are derived from the direct product of the monolayer symmetries, $\mathcal{M}_{x,y}^{\mathrm{mono}}$, expressed as $\mathcal{M}_{x,y}^{\mathrm{bi}}=\mathbb{I}_2 \otimes \mathcal{M}_{x,y}^{\mathrm{mono}}$.

The Wannier bands exhibit distinct structures below and above the critical interlayer coupling strength, $t_\mathrm{c}^{\rm{bilayer}}$. 
The nested Wilson loop calculation of the Wannier bands yields band polarizations of $p_y=\lbrace 1/2, 0, 0, 1/2 \rbrace$ for $t_\mathrm{v} < t_\mathrm{c}^{\rm{bilayer}}$ and $p_y=\lbrace 0, 0, 0, 0 \rbrace$ for $t_\mathrm{v} > t_\mathrm{c}^{\rm{bilayer}}$. 
This change in polarization, corresponding with the disappearance of corner states in finite systems, indicates a transition to a strong interlayer coupling regime, reducing the quantized quadrupole moment from one to zero.

The Wilson loop calculation applied to the occupied bands of the trilayer structure in Fig.~1(e), yields six non-degenerate Wannier bands (Fig.~\ref{fig:Wannierband}(b1,b2)). 
As $t_\mathrm{v}$ increases, the Wannier bands experience changes, leading to different polarizations. 
For $t_\mathrm{v} < t_\mathrm{c}^{\rm{trilayer}}$, we obtain the polarizations of $p_y = \lbrace 1/2, 1/2, 0, 0, 1/2, 1/2 \rbrace$, identifying the phase as trivial even with the existence of corner states. In contrast, for $t_\mathrm{v} > t_\mathrm{c}^{\rm{trilayer}}$, the polarizations are $p_y = \lbrace 1/2, 0, 0, 0, 0, 1/2 \rbrace$, indicating a topological phase.

For the alternative trilayer structure depicted in Fig.~1(f), the Wilson loop calculation generates six non-degenerate Wannier bands (Fig.~\ref{fig:Wannierband}(c1,c2)). 
The nested Wilson loop computation yields polarizations of $p_y = \lbrace 1/2, 0, 0, 0, 0, 1/2 \rbrace$ for $t_\mathrm{v} < t_\mathrm{c}^{\rm{trilayer}}$ and $p_y = \lbrace 0, 0, 0, 0, 0, 0 \rbrace$ for $t_\mathrm{v} > t_\mathrm{c}^{\rm{trilayer}}$. 
These polarization changes correspond to the number of corners obtained in finite-sized system calculations.

\section*{Appendix D: Perturbation theory}
To provide a clearer understanding of the results obtained from numerical simulations at large coupling strengths $t_{\text{v}}$, we have derived the effective Hamiltonian for these significant interlayer couplings. 
This derivation reveals both the presence and absence of four corner states within the topological heterostructures under large interlayer couplings.

($\mathrm{i}$)
For the bilayer system, we consider the interlayer coupling Hamitonian $H_{0}$, defined by
\begin{equation}
\mathcal{H}_{0}=\left( 
\begin{array}{cc}
0 & \mathcal{H}_{\text{v}}  \\ 
\mathcal{H}_{\text{v}} & 0  
\end{array}
\right) ,
\end{equation}
where only the interlayer couplings exist. 
This is diagonalized as 
\begin{equation}
U_{0}^{-1}\mathcal{H}_{0}U_{0}=\left( 
\begin{array}{cc}
-t_{\text{v}}\mathbb{I}_{4} & 0  \\ 
0 & t_{\text{v}}\mathbb{I}_{4}
\end{array}
\right) ,  
\label{UHUbi}
\end{equation}
where $\mathbb{I}_{4}$ is the four-by-four identity matrix.
The transformation matrix $U_{0}$ is given by
\begin{equation}
U_{0}=\left( 
\begin{array}{cc}
-\mathbb{I}_2 \otimes \sigma_x & \mathbb{I}_{4} \\ 
\mathbb{I}_2 \otimes \sigma_x & \mathbb{I}_{4}
\end{array}
\right) .
\end{equation}
When extracting either the upper or lower bands, such as extracting the upper band from equation (20) using the projection operator 
$P_{0}$, where $P_{0}$ projects onto the upper four-by-four submatrix, the resulting Hamiltonian for the upper four-by-four submatrix is given by 
\begin{widetext}
\begin{equation}
P_{0}^{-1}U_{0}^{-1}\mathcal{H}_{\text{bilayer}}U_{0}P_{0}=\left( 
\begin{array}{cccc}
t_{\text{v}} & 0 & \frac{1}{2}\left(1+e^{ik_{x}}\right)\left(\kappa _{1}+\kappa _{2}\right)  & \frac{1}{2}\left(1+e^{ik_{y}}\right) \left(\kappa _{1}+\kappa _{2}\right) \\ 
0 & t_{\text{v}} & \frac{1}{2}\left(1+e^{-ik_{y}}\right) \left(\kappa _{1}+\kappa _{2}\right)    & - \frac{1}{2}\left(1+e^{-ik_{x}}\right) \left(\kappa _{1}+\kappa _{2}\right)  \\ 
\frac{1}{2}\left(1+e^{-ik_{x}}\right) \left(\kappa _{1}+\kappa _{2}\right)  & \frac{1}{2}\left(1+e^{ik_{y}}\right) \left(\kappa _{1}+\kappa _{2}\right) & t_{\text{v}} & 0 \\ 
\frac{1}{2}\left(1+e^{-ik_{y}}\right) \left(\kappa _{1}+\kappa _{2}\right) & -\frac{1}{2}\left(1+e^{ik_{x}}\right) \left(\kappa _{1}+\kappa _{2}\right) & 0 & t_{\text{v}} 
\end{array}
\right) .
\end{equation}
\end{widetext}
At the M-point ($k_x=k_y=\pi$), all off-diagonal elements in Eq. (22) vanish. 
Consequently, equation (22) is always gapless at the M-point with an
energy offset of $t_{\text{v}}$ independent of the ratio of $\kappa_1$ and $\kappa_2$.
To further investigate the existence of localized states, we calculate the inverse participation ratio (IPR) and confirm the absence of localized corner states at all energy levels in the strong interlayer coupling regime of the bilayer system. The detailed calculations are provided in Appendix E.

($\mathrm{ii}$)
For the trilayer 'topological-trivial-topological' structure, we first diagonalize the interlayer coupling Hamiltonian $H_{0}$ defined by
\begin{equation}
\mathcal{H}_{0}=\left( 
\begin{array}{ccc}
0 & \mathcal{H}_{\text{v}} & 0 \\ 
\mathcal{H}_{\text{v}} & 0 & \mathcal{H}_{\text{v}} \\ 
0 & \mathcal{H}_{\text{v}} & 0
\end{array}
\right) ,
\end{equation}
where only the interlayer coupling exist. 
It is diagonalized as
\begin{equation}
U_{0}^{-1}\mathcal{H}_{0}U_{0}=\left( 
\begin{array}{ccc}
-\sqrt{2}t_{\text{v}}\mathbb{I}_{4} & 0 & 0 \\ 
0 & 0 & 0 \\ 
0 & 0 & \sqrt{2}t_{\text{v}}\mathbb{I}_{4}
\end{array}
\right) ,  \label{UHU}
\end{equation}
and the transformation matrix $U_{0}$ is given by
\begin{equation}
U_{0}=\left( 
\begin{array}{ccc}
\mathbb{I}_{4} & -\sqrt{2}\mathbb{I}_2 \otimes \sigma_x & \mathbb{I}_{4} \\ 
-\mathbb{I}_{4} & 0 & \mathbb{I}_{4} \\ 
\mathbb{I}_{4} & \sqrt{2}\mathbb{I}_2 \otimes \sigma_x & \mathbb{I}_{4}
\end{array}
\right) .
\end{equation}
We are interested in the middle four-by-four matrix, where there are three
degenerate zero energy states in equation (\ref{UHU}). 
Next, we transform the full Hamiltonian and take the middle four-by-four matrix, which is given by
\begin{widetext}
\begin{equation}
P_{0}^{-1}U_{0}^{-1}\mathcal{H}_{\text{trilayerI}}U_{0}P_{0}=\left( 
\begin{array}{cccc}
0 & 0 & \kappa _{1}+\kappa _{2}e^{-ik_{x}}  & \kappa _{1}+\kappa
_{2}e^{ik_{y}} \\ 
0 & 0 & \kappa _{1}+\kappa _{2}e^{-ik_{y}}  & - \left( \kappa _{1}+\kappa _{2}e^{ik_{x}} \right) \\ 
 \kappa _{1}+\kappa _{2}e^{ik_{x}}  & \kappa _{1}+\kappa_{2}e^{ik_{y}} & 0 & 0 \\ 
\kappa _{1}+\kappa _{2}e^{-ik_{y}} & -\left( \kappa _{1}+\kappa_{2}e^{-ik_{x}}\right) & 0 & 0 
\end{array}
\right) ,
\end{equation}
\end{widetext}
where $P_{0}$ is a projection operator taking the middle four-by-four
matrix. 
This results in an effective Hamiltonian, which is equivalent to the BBH model. 
Consequently, there are four corner modes for large interlayer couplings $t_{\text{v}}$ when $\kappa_1 < \kappa_2$. 
A similar discussion applies to the absence of corner modes for large interlayer couplings when $\kappa_1 > \kappa_2$, referred to as 'trivial-topological-trivial', as shown in Fig. 5(a). 
The detailed calculations are given in Appendix E.

\section*{Appendix E: Inverse participation ratio}
To investigate the presence of localized states, we calculate the inverse participation ratio (IPR) and confirmed the existence or absence of localized corner states across all energy in BBH bilayer and trilayer systems, both below and above the critical interlayer coupling strength where topological transitions occur.

The IPR quantifies the degree of localization of a wavefunction and is defined as
$\text{IPR}=\sum_i |\psi_i|^4$    
where $\psi_i$ represents the wavefunction at site $i$. 
A higher IPR value indicates stronger localization, while a lower value corresponds to a more delocalized state.

($\mathrm{i}$) BBH Bilayer System

\begin{figure}[t]
\includegraphics[]{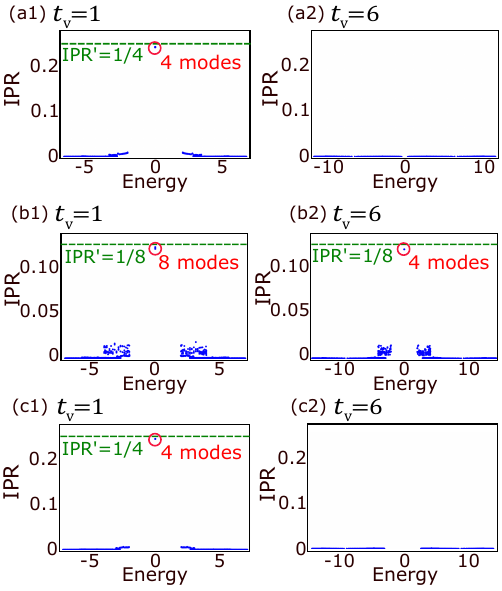}
\caption{\label{fig:IPR} 
IPR of (a) BBH bilayer,
(b) 'Topological-Trivial-Topological' BBH trilayer and 
(c) 'Trivial-Topological-Trivial' BBH trilayer systems.
(a1)-(c1) Below the critical interlayer coupling strength ($t_\mathrm{v} =1$).
(a2)-(c2) Above the critical interlayer coupling strength ($t_\mathrm{v} =6$).
Parameters are chosen as $\kappa_1=1$, $\kappa_2=3$, and $N$=20. 
The red circle indicates the IPR of the corner states, whereas the green dashed line indicates the ideal IPR value (IPR') when the corner states are entirely confined and evenly distributed among the corners.}
\end{figure}

When the interlayer coupling is weak, the IPR calculation shows four degenerate peaks with IPR$\simeq	$1/4 at energy $E=0$, indicating the presence of four degenerate corner states (Fig.~\ref{fig:IPR} (a1)). 
In particular, the IPR value should be 1/4 when each corner mode is predominantly localized at the four corners in layer 1, as given by $\text{IPR'}=4\times(1/16)=1/4$.
In the strong interlayer coupling regime, the maximum IPR is zero, indicating the absence of localized states (Fig.~\ref{fig:IPR} (a2)).

($\mathrm{ii}$) BBH trilayer system
type $\mathrm{I}$ (topological-trivial-topological structure)

In the weak interlayer coupling regime, there are eight modes with IPR $\simeq$ 1/8 at $E=0$, indicating the presence of eight localized states, each primarily confined to the eight corners in layers 1 and 3 (Fig.~\ref{fig:IPR} (b1)).
The ideal IPR value is 1/8 when each corner mode is localized at the eight corners across two layers, as given by $\text{IPR'} = 8 \times (1/64) = 1/8$.
As shown in Fig.~\ref{fig:IPR} (b2), the maximum IPR value remains unchanged in the strong interlayer coupling regime, but the number of corner states decreases to four as the corner modes arising from symmetrical couplings (Corner-sym) vanish.

($\mathrm{iii}$) BBH trilayer system
type $\mathrm{II}$ (trivial-topological-trivial structure)

In the weak interlayer coupling regime, there are four modes with IPR $\simeq$ 1/4 at $E=0$, indicating the presence of four degenerate corner states (Fig.~\ref{fig:IPR} (c1)). Here, the ideal IPR value is given by $\text{IPR'}=4\times(1/16)=1/4$.
When the system enters the strong interlayer coupling regime, the maximum IPR becomes zero, indicating the absence of localized states, as shown in Fig.~\ref{fig:IPR} (c2).

\section*{Appendix F: Chiral symmetric Hamiltonian in real space}
As an illustrative example, we provide a detailed calculation for $N=2$ for bilayer systems.
The submatrix of BBH bilayer Hamiltonian in equation (6) is 
composed of four 
$2N^2 \times 2N^2$ submatrices, described as
\begin{equation}
    h_{\text{bilayer}}=
    \begin{pmatrix}
h_{\text{Topo}}  & \mathcal{H}_{\text{vp}}\\
\mathcal{H}_{\text{vp}} & h_{\text{Tri}}
\end{pmatrix},
\end{equation}
where 
\begin{equation}
    h_{\text{Topo}}=
    \begin{pmatrix}
-\kappa_1 & 0 & 0 & 0 & \kappa_1 & 0 & 0 & 0\\
-\kappa_2 & -\kappa_1 & 0 & 0 & 0 & \kappa_1 & 0 & 0\\
0 & 0 & -\kappa_1 & 0 & \kappa_2 & 0 & \kappa_1 &  0 \\
0 & 0 & -\kappa_2 & -\kappa_1 & 0& \kappa_2 & 0 & \kappa_1 \\
\kappa_1 &  0 & \kappa_2 & 0&  \kappa_1  & \kappa_2 & 0 & 0 \\
0& \kappa_1 &  0 & \kappa_2 & 0 & \kappa_1 & 0 & 0\\
0 &  0 & \kappa_1 & 0& 0 & 0 & \kappa_1  & \kappa_2\\
0 & 0 &  0 & \kappa_1 & 0& 0 & 0 & \kappa_1
\end{pmatrix},
\end{equation}
and 
$\mathcal{H}_{\text{vp}}=t_{\text{v}} \times \mathbb{I}_{2N^2}$.
Note that $h_{\text{Topo}}$ and $h_{\text{Tri}}$ have an inverse relationship with respect to the ratio between $\kappa_1$ and $\kappa_2$.

The sublattice quadrupole moment operator $\mathcal{Q}_{xy}^{S}$ of the BBH bilayer system is a diagonal matrix where the diagonal elements form a repeating sequence of ($-i,-1,-1,1$), this sequence repeats $2N$ times.
Thus, $\mathcal{Q}_{xy}^{S}$ = $\mathbb{I}_4\otimes \text{diag.}(-i,-1,-1,1)$.

The calculation process for bilayer systems can be readily extended to trilayer systems.
The submatrix of BBH trilayer Hamiltonian is
\begin{equation}
    h_{\text{trilayer}}=
    \begin{pmatrix}
h_{\text{Topo}}  & \mathcal{H}_{\text{vp}} & 0\\
\mathcal{H}_{\text{vp}} & h_{\text{Tri}} & \mathcal{H}_{\text{vp}} \\
0 & \mathcal{H}_{\text{vp}} & h_{\text{Topo}}
\end{pmatrix}.
\end{equation}
The matrix $\mathcal{Q}_{xy}^{S}$ of the BBH trilayer system is given by $\mathcal{Q}_{xy}^{S}$ = $\mathbb{I}_6\otimes \text{diag.}(-i,-1,-1,1)$ for $N=2$.

\section*{Acknowledgement}
N.I. was supported by the JSPS KAKENHI (Grant No. JP21J40088) and CREST, JST (Grant 380 No. JPMJCR19T1). M.E. was supported by CREST, JST (Grants No. JPMJCR20T2) and by the Grants-in-Aid for Scientific Research from MEXT KAKENHI (Grant No.23H00171).
Y.O. was supported by the Grants-in-Aid for Scientific Research from MEXT KAKENHI (Grants No.22H01994 and No. 22H00298). 
S.I. was supported by CREST, JST (Grant No. JPMJCR19T1) and the Grants-in-Aid for Scientific Research from MEXT KAKENHI (Grants No 22H00298 and No. 22H01994).

\nocite{*}

\bibliography{BBHbilayer}

\end{document}